\documentstyle[12pt,epsfig]{article}
\def\cite#1{#1}

\def\thebibliography#1{\section*{References}\list
 {[\arabic{enumi}]}{\settowidth\labelwidth{[#1]}\leftmargin\labelwidth
 \advance\leftmargin\labelsep
 \usecounter{enumi}}
 \def\newblock{\hskip .11em plus .33em minus -.07em}
 \sloppy
 \sfcode`\.=1000\relax}

\title {Pseudoscalar meson transition form factors
\footnote{Contribution presented at the PHOTON'03, April 7-11,
2003, Frascati (Roma), Italy}}

\author{A.-Z.~Dubni\v{c}kov\'a$^1${\footnote{e-mail: dubnickova@fmph.uniba.sk}}, S.~Dubni\v{c}ka$^2$, G. Pancheri$^3$ and R. Pek\'arik$^2$}
\date{\empty}

\begin{document}
\maketitle
\begin{center} {
 $^{1}$ Dept. of Theor. Physics, Comenius Univ., Bratislava,
Slovak Republic \\$^{2}$ Inst. of Physics, Slovak Academy of
Sciences, Bratislava,
Slovak Republic \\
$^{3}$Lab. Nazionali di Frascati dell' INFN, via E. Fermi 40, I-0044 Frascati (Roma), Italy} \\
\end{center}

\vspace{0.2cm}

\begin{abstract}

Present-day experimental and phenomenological situation about
pseudoscalar ($\pi^0$, $\eta$, $\eta'$) meson transition form
factors is briefly reviewed. Necessity of more sophisticated
behaviours in time-like region of these form factors is
emphasized. Four-resonance unitary and analytic model of the
pseudoscalar meson transition form factors is constructed, which
describes well all existing experimental information.
 \vspace{1pc}
\end{abstract}

\maketitle

\section{INTRODUCTION}

Nowadays we know, that hadrons have a finite size, which in
electromagnetic (EM) interactions is revealed as the EM structure
of hadrons, phenomenologically to be described by one or more (it
depends from the spin of hadron under consideration) functions of
one variable (the four-momentum transfer squared $t=q^2=-Q^2$ of a
virtual photon), called the EM form factors (FF's). If in the
vertex with three lines appear one virtual photon and two
identical strongly interacting particles, one speaks about the
elastic EM FF's of hadron. If there is one virtual photon and two
different hadrons (or one hadron and one real photon) we speak
about the transition FF's. A subgroup of the latter represent the
pseudoscalar meson transition FF's, describing any $\gamma^*\to
\gamma P$ transition, where $P$ can be $\pi^0$, $\eta$ and
$\eta'$.

Unlike the elastic meson and baryon EM FF's, the pseudoscalar
meson transition FF's are not well elaborated till now, though
there are special physical problems in which they play an
important role. For instance, there is no evaluation of the
$e^+e^-\to \gamma P$ contributions to the muon anomalous magnetic
moment (in connection with its precise measurement at the E-821
experiment in BNL) just due to the absence of a reliable behaviour
of the corresponding transition FF's in the time-like region.

In this contribution we solve the latter problem by a construction
of unitary and analytic models of the pseudoscalar meson
transition FF's, which reproduce existing experimental data in
space-like and time-like region quite well.

\section{PRESENT THEORETICAL NOTIONS}

There is a single FF for each $\gamma^*\to \gamma P$ transition to
be defined by a parametrization of the matrix element of the EM
current $J_{\mu}^{EM}=2/3\bar u\gamma_{\mu} u-1/3 \bar d
\gamma_{\mu} d-1/3 \bar s \gamma s$
\begin{equation}
\langle P(p)\gamma(k)|J_{\mu}^{EM}|0\rangle =
\epsilon_{\mu\nu\alpha\beta}
p^{\nu}\epsilon^{\alpha}k^{\beta}F_{\gamma P}(q^2), \label{a1}
\end{equation}
where $\epsilon^{\alpha}$ is the polarization vector of the photon
$\gamma$,  $\epsilon_{\mu\nu\alpha\beta}$ appears as only the
pseudoscalar meson belongs to the abnormal spin-parity series.

A straightforward calculation of $F_{\gamma P}(Q^2)$ behaviour in
QCD is  impossible and one can obtain \cite{Brod81} in the
framework of PQCD only the asymptotic behaviour
\begin{equation}
lim_{Q^2\to \infty} Q^2F_{\gamma P}(Q^2)= 2f_P, \label{a2}
\end{equation}
where $f_P$ is the meson weak decay constant.

The behaviour of $F_{\gamma P}(Q^2)$ for $Q^2\to 0$ can be
determined from the axial anomaly in the chiral limit of QCD
\begin{equation}
lim_{Q^2\to 0} F_{\gamma P}(Q^2)= \frac{1}{4\pi^2f_P}\equiv
F_{\gamma P}(0). \label{a3}
\end{equation}

In order to describe the soft nonperturbative region of $Q^2$ a
simple interpolation formula has been proposed by Brodsky, Lepage
\cite{Brod81}
 \begin{equation}
F_{\gamma
P}(Q^2)=\frac{1}{4\pi^2f_P}\cdot\frac{1}{1+(Q^2/8\pi^2f^2_P)}
\label{a4}
\end{equation}
which, however,  does not describe even the space-like data well.
As a consequence, in the space-like region ($t< 0$) one usually
fits the experimentally observed $t=-Q^2$ dependence of $F_{\gamma
P}(Q^2)$ by a normalized empirical formula
\begin{equation}
F_{\gamma P}(t)/F_{\gamma
P}(0)=\frac{1}{(1-t/\Lambda_P)},\label{a5}
\end{equation}
where $1/\Lambda_P$=$\langle r^2_P \rangle/6$ is related to the
size $\langle r^2_P\rangle$ of the pseudoscalar meson P.

And in the time-like region ($t>0$),  of the well known
vector-meson resonance behaviour, one is left with Breit-Wigner
extension of the VMD model
\begin{equation}
F_{\gamma
P}(t)=\sum_v\frac{f_{vP\gamma}}{f_v}\cdot\frac{m_v^2}{m_v^2-t-im_v\Gamma_v},
\label{a6}
\end{equation}
which, however, is justified only at the region of resonances and
generally violates the unitarity condition and the normalization.

From all mentioned above it follows that it is highly desirable to
construct more sophisticated model of the pseudoscalar meson
transition FF's.

\section{EXPERIMENTAL SITUATION}

There are dominating experimental data on the pseudoscalar meson
transition FF's in the space-like region
\cite{Berg84,Ber91,Aha90,Gron98}. However, recent measurements of
the $e^+e^-\to \eta \gamma$, $\pi^0\gamma$ processes at VEPP-2M in
Novosibirsk \cite{Acha99} around $\phi$-meson supplemented older
data \cite{Die80,Druz84,Dol89,Vik80} at lower energies and they
all provide a possibility to carry out a serious analysis of the
corresponding FF's. So, at present day we have all together
\begin{itemize}
\item for $\pi^0$: $\quad$ 33 points
\item for $\eta$: $\quad$ 52 points
\item for $\eta'$: $\quad$ 59 points.
\end{itemize}

In the next section we construct unitary and analytic model of the
pseudoscalar meson transition FF's which unifies all known
properties of FF's and the existing data will be utilized for an
evaluation of free parameters of the model.

\section{UNITARY AND ANALYTIC MODEL}

So, our intention is to achieve a description of all $t<0$ and
$t>0$ data on the pseudoscalar meson transition FF's by one,
however, distinct for $\pi^0$, $\eta$ and $\eta'$, analytic
function explicitly known on the real axis of $t$-plane from
$-\infty$ to $+\infty$, respecting all properties of $F_{\gamma
P}(t)$ like the asymptotic behaviour (\ref{a2}), the normalization
(\ref{a3}), the analytic properties of $F_{\gamma P}(t)$, the
reality condition $F_{\gamma P}^*(t)$=$F_{\gamma P}(t^*)$ from
which it directly follows a property of the unitarity condition,
that the $Im F_{\gamma P}(t)$ is $\neq 0$ only from the lowest
branch point on the positive real axis of $t$-plane to $+\infty$,
etc.

The transition FF $F_{\gamma P}(t)$ is suitable to split into two
terms depending on the isotopic character of the photon
\begin{equation}
F_{\gamma P}(t)=F_{\gamma P}^{I=0}(t)+ F_{\gamma
P}^{I=1}(t),\label{a7} \end{equation} where $F_{\gamma
P}^{I=0}(t)$ can be saturated only by isoscalar vector mesons and
$F_{\gamma P}(t)^{I=1}$ can be saturated only by isovector
vector-mesons, whereby both sets are characterized by the photon
quantum numbers.

The analytic properties of $F_{\gamma P}(t)$ consist in the
assumption, that $F_{\gamma P}(t)$ is analytic in the  whole
complex $t$-plane besides the cut on the positive real axis from
$t_0=m_{\pi^0}^2$ up to $+\infty$, because there is the
intermediate $\pi^0\gamma$ state allowed in the unitarity
condition of every $\pi^0$, $\eta$ and $\eta'$ transition FF,
which generates just the lowest branch point $t_0=m_{\pi^0}^2$.
Moreover, from the unitarity condition it follows that there is an
infinite number of higher branch points on the positive real axis
of $t$-plane as there is allowed infinite number of higher
intermediate states in the unitarity condition of FF's under
consideration. In our model we restrict ourselves to two
square-root cut approximation of the latter picture. In order to
fuse such analytic properties into our model we start with the
following 4 resonance VMD model parametrization of (\ref{a7})
\begin{equation}
F_{\gamma
P}(t)=\sum_{s=\omega,\phi,\omega'}\frac{m_s^2}{m_s^2-t}(f_{s\gamma
P}/f_s^e)+ 
 \frac{m_{\rho}^2}{m_{\rho}^2-t}(f_{{\rho}\gamma
P}/f_{\rho}^e), \label{a8}
\end{equation}
and apply the normalization condition (\ref{a3}). However, in
order to take into account the fact, that $f_{\eta}$ and
$f_{\eta'}$ (unlike $f_{\pi}$) are not directly measurable
quantities, employing the relation for two-photon partial width
\begin{equation}
\Gamma(P\to \gamma\gamma)=\frac{\alpha^2}{64\pi^3f^2_P}m_P^3
\label{a9}
\end{equation}
of the pseudoscalar meson $P$, one comes to a redefinition of the
norm through the known partial two-photon width $\Gamma(P\to
\gamma\gamma)$ in the following way
\begin{equation}
F_{\gamma P}(0)=\frac{2}{\alpha m_P}\sqrt{\frac{\Gamma(P\to
\gamma\gamma)}{\pi m_P}}. \label{a10}
\end{equation}
Then from (\ref{a8}) one obtains the following condition on
coupling constant ratios
\begin{equation}
F_{\gamma P}(0)=\sum_{s=\omega,\phi,\omega'}(f_{s\gamma P}/f_s)+
(f_{{\rho}\gamma P}/f_{\rho}), \label{a11}
\end{equation}
from which one can express e.g. $(f_{\omega'\gamma
P}/f_{\omega'})$ by all others. Substituting the latter into
(\ref{a8}) one gets the VMD parametrization
\begin{eqnarray}
&&F_{\gamma P}(t)=F_{\gamma
P}(0)\frac{m_{\omega'}^2}{m_{\omega'}^2-t}+ \label{a12}
\\ &+& \left
[\frac{m_{\omega}^2}{m_{\omega}^2-t}-\frac{m_{\omega'}^2}{m_{\omega'}^2-t}
\right ]a_{\omega}+ \nonumber \\
&+& \left [
\frac{m_{\phi}^2}{m_{\phi}^2-t}-\frac{m_{\omega'}^2}{m_{\omega'}^2-t}\right
]a_{\phi}+\nonumber \\
&+& \left
[\frac{m_{\rho}^2}{m_{\rho}^2-t}-\frac{m_{\omega'}^2}{m_{\omega'}^2-t}\right
]a_{\rho}, \quad a_i=(f_{i\gamma P}/f_i)   \nonumber
\end{eqnarray}
to be normalized automatically. In order to obtain from
(\ref{a12}) the unitary and analytic  model, we incorporate the
two-cut approximation of the true analytic properties by an
application of the nonlinear transformation
\begin{equation}
t=t_0 -\frac{4(t_{in}^s-t_0)}{[1/V-V]^2};\quad  t=t_0
-\frac{4(t_{in}^v-t_0)}{[1/W-W]^2}  \label{a13}
\end{equation}
to the isoscalar and isovector ($\rho$-meson) terms and
subsequently also nonzero values  of vector-meson widths,
$\Gamma_s \neq 0$ and $\Gamma_{\rho} \neq 0$, are established. The
effective square-root branch points $t_{in}^s$ and $t_{in}^v$
include in average contributions of all higher important
thresholds in both, isoscalar and isovector case, respectively,
and are left to be free parameters of the constructed model. The
variable $V$ ($W$) in (\ref{a13}) is conformal mapping
\begin{equation}
V(t)=i\frac {\sqrt{q_{in}^{s}+q}-
 \sqrt{q_{in}^{s}-q}}
{\sqrt{q_{in}^{s}+q} + \sqrt{q_{in}^{s}-q}} \label{a14}
\end{equation}
$$
q=[(t-t_0)/t_0]; \quad  q_{in}^s=[(t_{in}^s-t_0)/t_0]$$ of the
four -sheeted Riemann surface in $t$-variable into one $V$-plane (
$W$-plane).

As a result of application of (\ref{a13}) all VMD terms in
(\ref{a12}) first give the factorized forms
\begin{eqnarray}
& &\frac{m^2_{i}}{(m^2_{i}-t)}=\left (\frac{1-V^2}{1-V_N^2} \right )^2\cdot \label{a15}\\
& &\frac{(V_N-V_{i_0})(V_N+V_{i_0})(V_N-1/V_{i_0})(V_N+1/V_{i_0})}
{(V-V_{i_0})(V+V_{i_0})(V-1/V_{i_0})(V+1/V_{i_0})}\nonumber
\end{eqnarray}
 on the pure asymptotic term $(\frac{1-V^2}{1-V_N^2})^2$,
independent on the flavour of vector-mesons under consideration
(however it depends on the isospin) and carrying just the
asymptotic behaviour $_{|t|\to\infty} \sim t^{-1}$ of the VMD
model, and the so-called resonant term (the second one in
(\ref{a15})) describing the resonant structure of VMD terms, which
however, for $|t|\to\infty$ is going out on the real constant and
so, it doesn't contribute to the asymptotic behaviour of the VMD
model

  The subindex 0 in (\ref{a15}) means that still $\Gamma$=0 of all
  vector mesons is considered.

  In order to demonstrate the reality condition $F_{\gamma P}^*(t)$= $F_{\gamma P}(t^*)$
  explicitly, one can utilize in (\ref{a15}) relations between
  complex conjugate values of the corresponding zero-width VMD
  model pole positions in $V$ (or $W$) plane
  \begin{equation}
V_{\omega_0}=-V_{\omega_0}^*; \quad
W_{\rho_0}=-W_{\rho_0}^*;\label{a16}
\end{equation}
 and
 $$ V_{i_0}=1/V_{i_0}^* \quad {\rm for}\quad
i=\phi,\omega'$$  following from the reality that in a fitting
procedure of existing data on $F_{\gamma P}(t)$ such $t_{in}^s$
(or $t_{in}^v$) is found that
\begin{equation}
(m_i^2-\Gamma_i^2/4)< t_{in}^s,t_{in}^v \quad i=\omega,\rho
\label{a17}
\end{equation}
 and
 $$ (m_j^2- \Gamma^2_j/4) > t_{in}^s \quad j=\phi, \omega'.$$

Finally, incorporating $\Gamma \neq 0$ by a substitution
\begin{equation}
m^2_r\rightarrow (m_r-i\Gamma_r/2)^2 \label{a18}
\end{equation}
one comes to the unitary and analytic  models of $F_{\gamma P}(t)$
in the following form
\begin{eqnarray}
&&F_{\gamma P}[V(t)]= \left (\frac{1-V^2}{1-V_N^2}\right )^2\cdot
\label{a19}\\
&\cdot& \left \{F_{\gamma P}(0){\bf {V}}_{\omega'}+ \left
[{\bf{V}}_{\omega}-{\bf{V}}_{\omega'}\right.\right ]a_{\omega}
+\nonumber
\\
&+&\left.\left [{\bf{V}}_{\phi}-{\bf{V}}_{\omega'}\right ]a_{\phi}-{\bf{V}}_{\omega'}a_{\rho}\right \}+ \nonumber \\
&+& \left
(\frac{1-W^2}{1-W_N^2}\right)^2{\bf{W}}_{\rho}a_{\rho}\nonumber
\end{eqnarray}
 where
$$
{\bf{V}}_{\bf{\omega'}}=
\frac{(V_N-V_{\omega'})(V_N-V_{\omega'}^*)(V_N+V_{\omega'})(V_N+V_{\omega'}^*)}{(V-V_{\omega'})(V-V_{\omega'}^*)(V+V_{\omega'})(V+V_{\omega'}^*)}
$$
$$
{\bf{V}}_{\bf{\omega}}=\frac{(V_N-V_{\omega})(V_N-V_{\omega}^*)(V_N-1/V_{\omega})(V_N-1/V_{\omega}^*)}{(V-V_{\omega})(V-V_{\omega}^*)(V-1/V_{\omega})(V-1/V_{\omega'}^*)}
$$
$$
{\bf{V}}_{\bf{\phi}}=\frac{(V_N-V_{\phi})(V_N-V_{\phi}^*)(V_N+V_{\phi})(V_N+V_{\phi}^*)}{(V-V_{\phi})(V-V_{\phi}^*)(V
+V_{\phi})(V+V_{\phi}^*)}
$$
$$
{\bf{W}}_{\bf{\rho}}=\frac{(W_N-W_{\rho})(W_N-W_{\rho}^*)(W_N-1/W_{\rho})(W_N-1/W_{\rho^*})}{(W-W_{\rho})(W-W_{\rho}^*)
(W-1/W_{\rho})(W-1/W_{\rho}^*)}
$$
Now, substituting instead of $P$ in (\ref{a19}) $\pi^0$, $\eta$
and $\eta'$, we get three different unitary and analytic models
for $F_{\gamma \pi^0}(t)$, $F_{\gamma \eta}(t)$ and $F_{\gamma
\eta'}(t)$ respectively. They are applied to a description of
existing data on $\pi^0$, $\eta$ and $\eta'$ transition FF's.

\begin{figure}[htp] 
\centering
\includegraphics[scale=.4
]{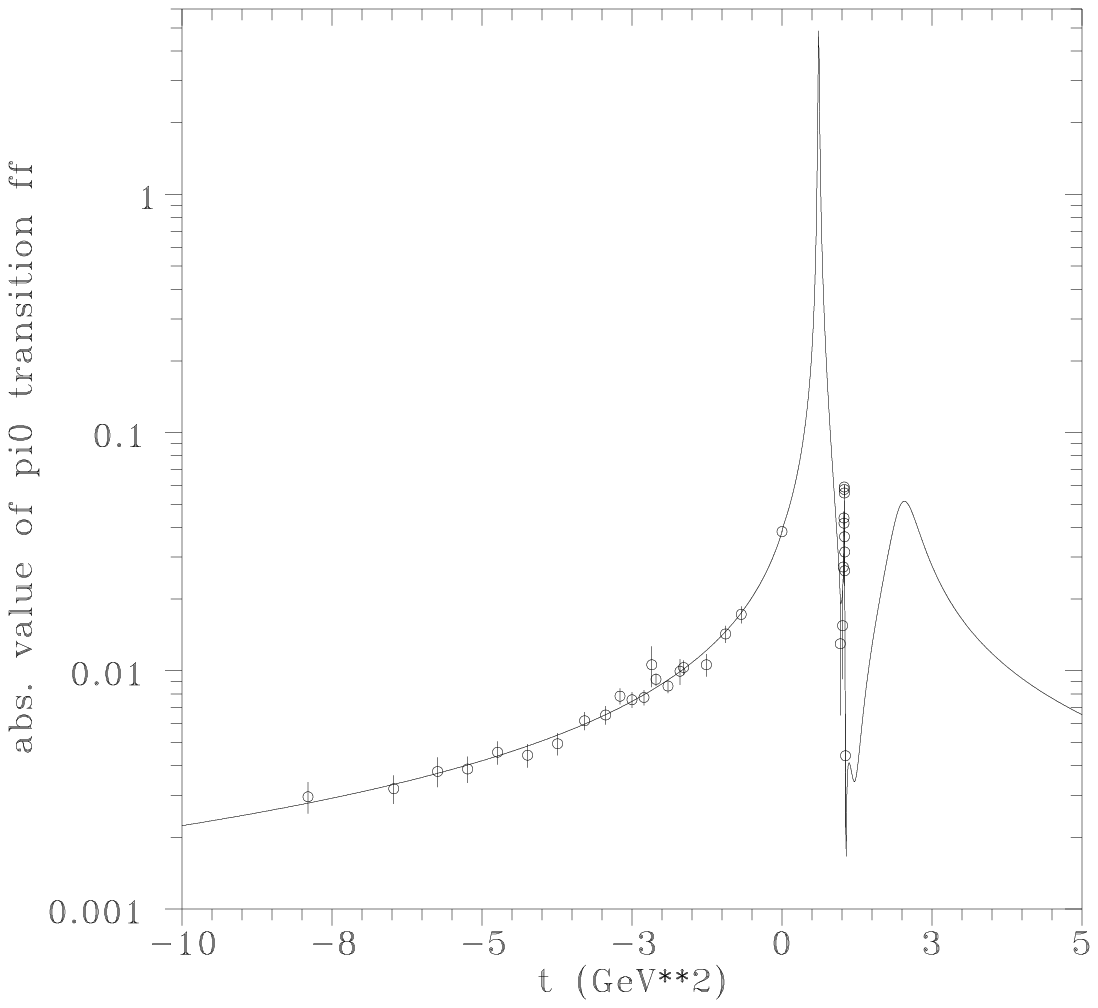} \caption{$\pi^0$ transition form
factor.}\label{fig:1}
\end{figure}

 $\pi^0$: The best description (see Fig.1) is achieved with
$\chi^2$ =23.2 i.e. $\chi^2/ndf$ =0.83 and following values of
parameters :

$t_{in}^s=0.6081 GeV^2; \quad t_{in}^v=1.0464 GeV^2$\\
$(f_{\omega\gamma\pi^0}/f_{\omega})$=$0.1129\pm 0.0007$\\
$(f_{\phi\gamma\pi^0}/f_{\phi})$=$-0.0002\pm 0.0001$\\
$(f_{\rho\gamma\pi^0}/f_{\rho})$=$-0.0769\pm 0.0007$ \\

$\eta$: The best description (see Fig.2) is achieved with $\chi^2$
=47.5 i.e. $\chi^2/ndf$ =1.01 and following values of parameters :
$t_{in}^s=1.0323 GeV^2; \quad t_{in}^v=1.5414 GeV^2$\\
$(f_{\omega\gamma\eta}/f_{\omega})$=$-.0524\pm 0.0201$\\
$(f_{\phi\gamma\eta}/f_{\phi})$=$-0.0013\pm 0.0001$\\
$(f_{\rho\gamma\eta}/f_{\rho})$=$0.0821\pm 0.0202$

\begin{figure}[htp] 
\centering
\includegraphics[scale=.4
]{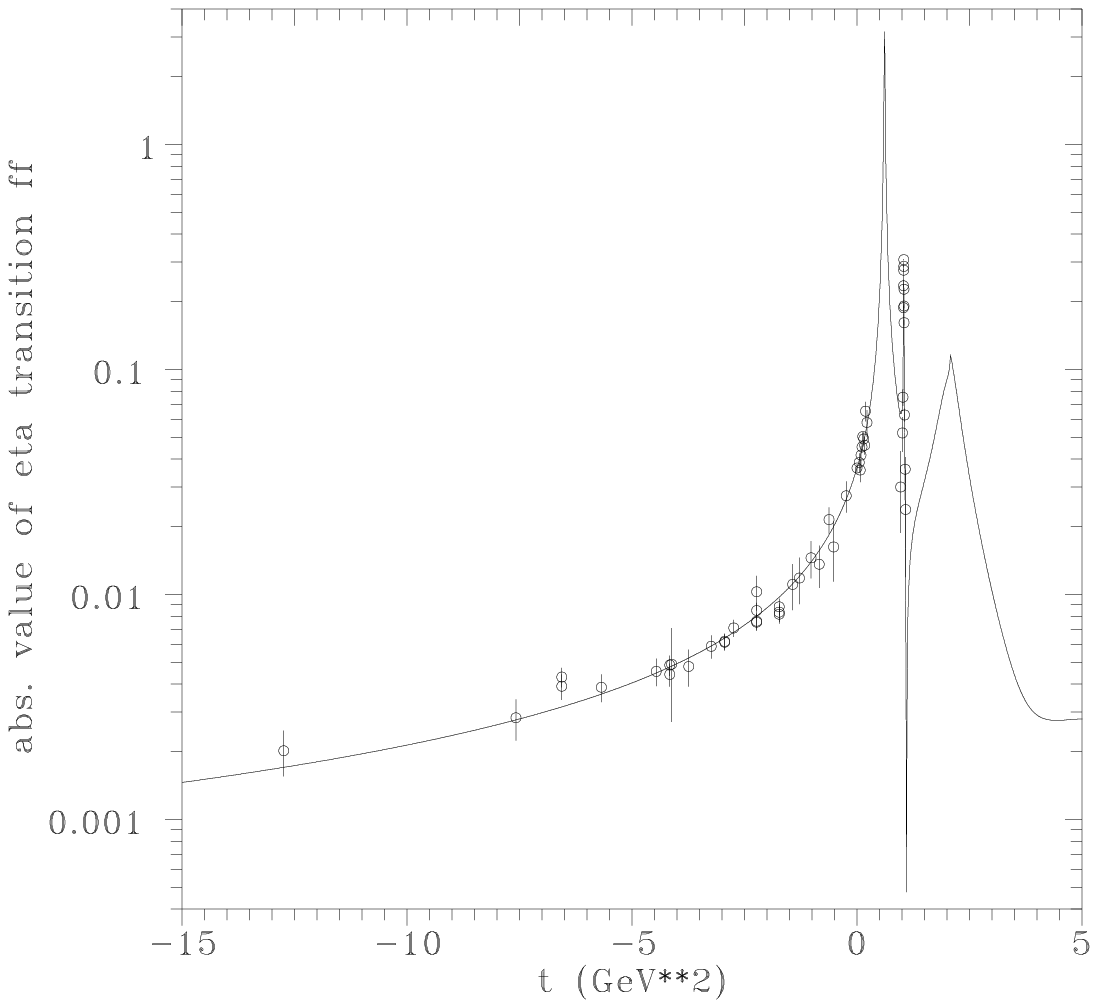} \caption{$\eta$ transition form
factor.}\label{fig:2}
\end{figure}

\begin{figure}[htp]
\centering
\includegraphics[scale=.4
]{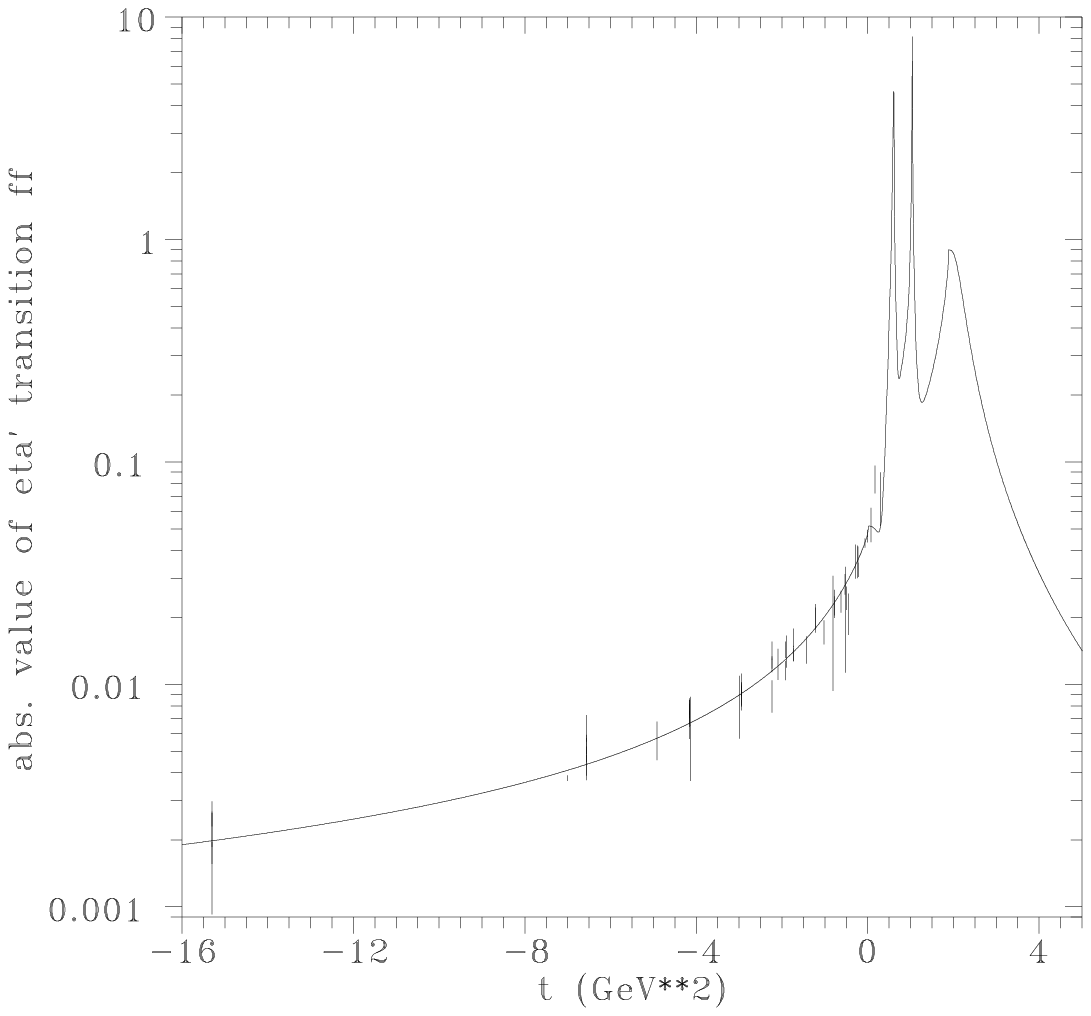} \caption{$\eta'$ transition form
factor.}\label{fig:3}
\end{figure}

  $\eta'$: The best description (see Fig.3) is achieved with
$\chi^2$ =67.4 i.e. $\chi^2/ndf$ =1.25 and following values of
parameters :

$t_{in}^s=1.0374 GeV^2; \quad t_{in}^v=1.5174 GeV^2$\\
$(f_{\omega\gamma\eta'}/f_{\omega})$=$-.1623\pm 0.0020$\\
$(f_{\phi\gamma\eta'}/f_{\phi})$=$0.1182\pm 0.0052$\\
$(f_{\rho\gamma\eta'}/f_{\rho})$=$0.1253\pm 0.0013$

\section{CONCLUSIONS}

We have briefly reviewed the present-day experimental and
theoretical situation about the pseudoscalar meson transition FF's
and came to the conclusion that there is no reliable prediction of
them in the time-like region. Since there are besides quite good
experimental information in the space-like region new Novosibirsk
measurements of $e^+e^-$ annihilation processes into $P \gamma$ at
the $\phi$-meson peak, we have constructed the well founded
unitary and analytic model of $F_{\gamma \pi^0}(t)$, $F_{\gamma
\eta}(t)$ and $F_{\gamma \eta'}(t)$ with four vector-mesons and
predicted a reasonable behaviour of these transition FF's in the
space-like and time-like regions simultaneously. Results can be
applied for a prediction of various cross-sections and decay rates
in which pseudoscalar meson transition FF's appear, further, an
evaluation of contributions of $e^+e^-\to \gamma P$ processes into
the muon anomalous magnetic moment can be carried out and also the
strange pseudoscalar meson transition FF's can be in principle
predicted for the first time.

This work was in part supported by Slovak Grant Agency for
Sciences, Grant 2/1111/22 (S.D. and  R.P.) and Grant 1/7068/22
(A.Z.D.)


\begin{thebibliography}{9}
\bibitem{Brod81} S.J. Brodsky and G.P. Lepage, Phys. Rev. D 24  (1981) 1808.
\bibitem{Berg84} Ch. Berger et al, Phys. Lett. 142 B (1984) 125.
\bibitem{Ber91} H.-J. Behrend et al, Z. Phys. C49 (1991) 401.
\bibitem{Aha90} H. Aihara et al, Phys. Rev. Lett. 64 (1990) 172.
\bibitem{Gron98} J. Gronberg et al, Phys. Rev. D57 (1998) 33.
\bibitem{Acha99} M.N. Achasov et al, preprint Budker INP 99-39,
                 Novosibirsk (1999).
\bibitem{Die80} R.I. Djhelyadin et al, Phys. Lett. 94B (1980) 548.
\bibitem{Druz84} V.P. Druzhinin et al, Phys. Lett. 144B (1984) 136.
\bibitem{Dol89} S.I. Dolinsky et al, Z. Phys C42 (1989) 511.
\bibitem{Vik80} V.A. Viktorov et al, Sov. J. Nucl. Phys. 32
                (1980) 520.
\end{thebibliography}
\end{document}